\documentclass{article}

\usepackage{PRIMEarxiv}
\usepackage{natbib}
\usepackage[utf8]{inputenc}
\usepackage[T1]{fontenc}
\usepackage{hyperref}
\usepackage{url} 
\usepackage{booktabs}
\usepackage{amsfonts}
\usepackage{amsmath}
\usepackage{nicefrac}
\usepackage{microtype}
\usepackage{lipsum}
\usepackage{fancyhdr}
\usepackage{graphicx}       
\graphicspath{{media/}}     

\title{The Need for Quantitative Resilience Models and Metrics in Classical-Quantum Computing Systems}

\author{
  Santiago Núñez-Corrales \\
  National Center for Supercomputing Applications \\
  University of Illinois Urbana-Champaign \\
  Urbana IL\\
  \texttt{nunezco2@illinois.edu}
}

\begin{document}
\maketitle

\begin{abstract}
Increasingly deeper integration of HPC resources and QPUs unveils new challenges in computer architecture and engineering. As a consequence, dependability arises again as a concern encompassing resilience, reproducibility and security. The  properties of quantum computing systems involve a reinterpretation of these factors in retrodictive, predictive, and prescriptive ways. We state here that resilience must become an \emph{a priori} design constraint rather than an afterthought of HPC-QPU integration. This article describes the need for conceptual and quantitative models to estimate and assess the resilience hybrid classical-quantum computing infrastructure. We suggest how resilience methods in civil engineering can apply at various levels of the classical-quantum computing stack. We also discuss implications of a model of end-user value for the estimation of consequences resulting from the propagation of vulnerabilities from a given level of the stack upwards. Finally, we argue in favor of new resilience models can help the impact of improving specific components in quantum technology stacks to provide a clearer picture about the value of separation of concerns across different layers. Ultimately, HPC-QPU integration will increasingly demand more precise statements about the cost-benefit ratio of specific system improvements and their cascading consequences against estimates of delivered value to users.
\end{abstract}


\section{Introduction}

The evolution of quantum computing hardware during the past decade from \textit{experimental testbeds} to \textit{systems} capable of demonstrating algorithmic promise involves interaction with classical systems on various fronts. At the onset, quantum hardware platforms require classical computing systems to translate discrete digital specifications of pulses into analog signals which control instrumentation to achieve delicate metastable states \citep{berberich2024quantum}. In a recapitulation of classical hardware development, manual tasks performed by skilled experimentalists with relatively limited accuracy, reproducibility, and scalability are increasingly performed by application-specific integrated circuitry (ASIC) as a pathway to identify which elements of quantum stacks should remain flexible and which ones should stabilize materially \citep{butko2020understanding}. At the same time, we remain at a point in which the standard quantum building block analogous to the transistor remains to be found, and will likely remain so for the next decade: multiple qubit modalities span the current landscape, each with various strengths and limitations \cite{chohan2024comparative}. The challenges involved in realizing qubits across distinct modalities as well as the need to avoid overcommitment with certain research and development pathways introduce tensions when transforming quantum platforms into proper computing systems.

More generally, quantum computers require general classical computing systems to deliver \textit{value} to their users \citep{gambetta2022quantum}. Current understanding of quantum algorithms and quantum computational complexity suggests that an increasing number of applied problems constitute prime targets for quantum processing units (QPUs). From unstructured (polynomial speedups) to structured (exponential speedups) \cite{aaronson2009need}, the set of problems quantum computers are expected to excel at intersects largely with scientific, industrial, and societal challenges at the top of our civilizational attention \cite{arnault2024typology} provided we accomplish the feat of integrating quantum hardware at scales of practical utility and, we hope, advantage. As a consequence of quantum mechanics, our high-level hardware interaction with QPUs from the perspective of program synthesis is restricted to state initialization, execution of pure quantum programs, and measurement of result states. Conceptually encapsulating the classical control system as part of the QPU results in a piece of computing hardware whose functionality and operation are no different than an FPU or a GPU before integration into a computer architecture. Classical information specifies inputs to be encoded for quantum programsas well as the interpretable outcome of quantum computations after ensemble measurements occur. Historically, the problems we intend to solve with quantum resources in a fundamentally faster way have been addressed by High-Performance Computing (HPC) through hardware-software co-design. HPC is justified when the computational complexity of a problem makes it resistant to either efficient or exact solution, and the only way forward is parallelization. Thus, HPC-QPU integration should come as no surprise \cite{beck2024integrating,elsharkawy2024integration}, with quantum kernels replacing parts of (usually parallel or distributed) programs that usually solve intractable problems.

In this article, we focus on \textit{resilience} for dependable classical-quantum computing \textit{systems} (DCQCS). \textit{DCQCS cyberinfrastructure} means here \textit{the collection of hardware, software, and network systems with classical and quantum parts whose integration leads to repeated access, use, and ultimately benefit}. We first describe reliability in the context of \textit{dependability}, a top-level objective in DCQCS engineering \citep{giusto2024dependable}. We then argue that \textit{resilience} captures the concerns related to the structural and operational properties of these systems much more closely than \textit{reliability}. We show how concepts from complex multiscale stochastic systems (CMSS) theory, compositionality, and dynamical systems accurately characterize reliability in the context of DCQCS. To operationalize these concepts effectively, we turn our attention to methods in civil engineering and their translation to classical-quantum computing systems. We suggest two paradigmatic examples of perturbations to quantum systems and how they fit the methods for the purposes of illustration. Finally, we address the need to integrate resilience into a larger scheme of incentives, which requires defining the value of compute systems more broadly. This article concludes with a call to action for the broader spectrum of scientific and engineering communities capable of conducting research and development to fully materialize the many aspects of resilience in DCQCS.

\section{Defining Resilience in Classical-Quantum Computer Systems}

We build computers because they are inherently useful devices; their problem-solving throughput, speed, and accuracy are orders of magnitude above that of humans. To remain beneficial, these systems must become \textit{dependable}. DCQCS should not be the exception \cite{giusto2024dependable}: on the contrary, the overarching theme of this manuscript is the need for better dependability theory and tools posed by the large number of emerging challenges during HPC-QPU integration in terms of resilience, reproducibility, and security and privacy (Figure \ref{fig:1}). Establishing the dependability of a system in which quantum computing resources interact with classical ones appears to require new intellectual machinery .

\subsection{Resilience is an aspect of dependability}

\begin{figure}[h!]
\begin{center}
\includegraphics[width=0.6\textwidth]{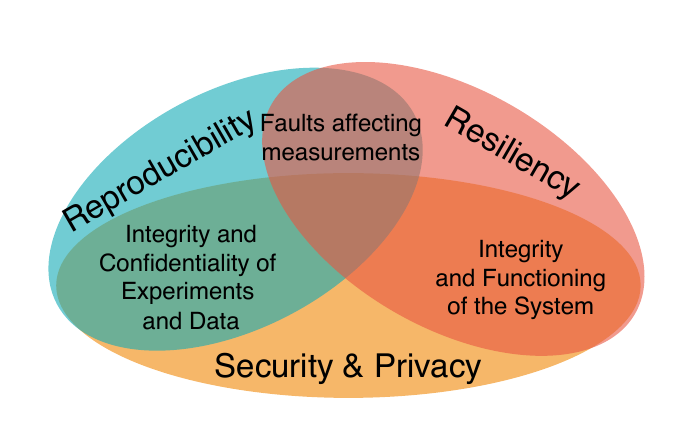}
\end{center}
\caption{Intersecting aspects of dependability in Classical-Quantum Computer Systems. Reproduced from \cite{giusto2024dependable}.}\label{fig:1}
\end{figure}

\textit{Dependablity} most generally expresses ``\textit{the ability to deliver service that can justifiably be trusted}'' \cite{avizienis2001fundamental}. We will later argue that \emph{trust} defines an aspect of the \emph{value} of a computer system which is hard to quantify. We increasingly want services that provide access to classical-quantum functionality characterized by a combined \emph{state} that undergoes a well-defined \emph{transitions}; in short, a DCQCS equates to an abstract machine of sorts. Such state transitions should be separable into those where the service produces intended behavior (i.e., \emph{correct service}) versus those in which deviations from it exist (i.e., \emph{incorrect service}). We will focus here on the intent behind assessing differences between expected and actual behavior of DCQCS (Figure \ref{fig:1}). Furthermore, a specific DCQCS provides a \textit{service interface} through mechanisms that allow users to access resources and place requests (i.e., \textit{demands}) on the system. There requests constitute stimuli with a formal specification for the purposes of dependability analysis. The concept of a user encompasses human, physical, and digital agents that consume services using its interface. When a local deviation from intended behavior occurs (i.e., a \emph{error}), we often attempt to trace its manifestation to a root cause (i.e., one or more \emph{faults}) since these tend to interact with the system and produce error cascades leading to incorrect service (i.e., \emph{failures}) in various degrees of severity. We find this language usually associated with resilience of all classes of computing infrastructure.

We naturally expect dependability --and in particular, resilience- aspects of quantum computing hardware by itself to be substantially more demanding than those of classical computers. At the foundations, we dynamically coerce easily perturbable stochastic metastable systems into useful quantum regimes achievable only within a narrow volume of tunable parameter space \citep{nielsen2006optimal,boscain2021introduction}; contrast this against the robustness of bistable systems with large built-in tolerances and at this point no parameters in the world of classical hardware. In terms of actual hardware platforms, optimal quantum control mixes with the specific hardware engineering challenges imposed by qubit modalities \citep{koch2022quantum}, increasing the number of moving parts in the system engineers must keep track of. We also lack quantum abstract machines at a sufficiently high level of expressiveness --well above quantum circuits \citep{nunez2023quantum}, which results in substantial limitations in the specification of quantum programs available to us today \citep{di2024abstraction} and ultimately in a general lack of ability to trace effectively and efficiently runtime aspects of dependability. The integration of HPC-QPU will remain challenging, as the rate of change in hardware and software greatly differs in both; this is effectively a \emph{stiff system} where frequencies of improvement differ across technologies and therefore co-design is essential. To summarize: DCQCS belong to the category of complex systems and are thus determined by Asbhy's law of requisite variety in terms of how much additional complexity is required to control them \cite{klir1991requisite}.

\subsection{Resilience applies naturally to DCQCS}

We will concentrate exclusively on resilience in DCQCS from here onward. By \emph{resilience}, we mean here the ability of a system to \emph{either maintain correct service, restore it, or degrade service quality as gracefully as possible by minimizing the probability of occurrence and impact of failures given certain stimuli associated with a non-zero probability of errors given both empirically known and theoretically estimated faults}. When applied to DCQCS, this involves theory, tools, and practice in multiple contexts. It also brings along a multitude of methods already in store for classical computing systems, since QPUs tend to integrate most parsimoniously as coprocessors or hardware accelerators. While HPC-QPU integration still must negotiate between multiple qubit modalities and degrees of proximity to HPC resources \cite{rallis2025interfacing}, we must foresee and plan for a time in which the integration happens closer, with much less software complexity and more rationally driven. 

The properties and challenges of quantum computing testbeds and quantum vendor hardware both diverge and converge depending on context in relation to integration efforts. Testbeds afford us direct visibility of all elements of the quantum stack from the hardware up to limited execution of quantum algorithms; these serve quantum information science and technology (QISE) research needs, not applications. Quantum vendor hardware at the other extreme seeks to materialize the promises of fundamental algorithmic speedups in service of the specific needs of application use cases. It is substantially different to perform a sequence of unique experiments (or their replication) related to a certain experimental hypothesis than to ensure a service-level agreement for routine paid use. The first case speaks of bespoke instrumentation where the stakes comprise producing new science advances, many of which are painstaking to achieve. The second case implies small margins of operational error, with repeatable access mechanisms that ensure reproducible behavior to deliver value that results in return on investment.

The main difference between both is the definition of what value means for their users: in the first case, value equals new discoveries in QISE; in the second, applied impacts and profit. The point of convergence is the effect resilience has on the impact of outcomes in both quantum testbeds and quantum vendor hardware. Reducing the challenges in QISE laboratories requires qubit platforms where practices encoded as standard operating procedures work resiliently and errors are easily traced when these occur. Quantum vendor hardware will be subject to a wide landscape of stimuli, some of which will be unexpected and potentially dangerous to preserving the quality of service --and the service interface itself- of the system. Moreover, an emphasis on resilience across both quantum computing testbeds and quantum vendor hardware will necessarily have a positive impact on the speed at which quantum computing systems evolve, the transferrable abilities from academia to industry regarding workforce training, and the likelihood of encountering and overcoming systemic barriers to scalability and fault tolerance.

These differences translate, ultimately, into overarching goals for HPC-QPU integration. Quantum testbeds when combined with HPC facilities result in reference architectures that expose all details involved in the process at a scale consumable for research and training purposes. Ongoing efforts at the National Center for Supercomputing Applications at UIUC with the Leadership Class Compute Facility project\footnote{See: \url{https://github.com/lccf-qhpc}.}, the Munich Quantum Valley \cite{burgholzer2026munich}, ORNL \cite{shehata2025building} and others recognize the need for such reference architectures to reduce future frictions for HPC facilities, as well as for the training of future generation of quantum computer engineers. The argument for quantum vendor systems stems from the need to understand the functional, physical and integration requirements into HPC facilities, many of which are either non-trivial (e.g., deep telemetry \cite{ahmed2025telemetry}) or still unknown. We argue here that, in both cases, the initial cost of designing under a DCQCS philosophy amortizes as a function of the resulting inspectability and reliability of elements of the system, enabling thus opportunistic refinement between service interfaces.

\begin{figure}[h!]
\begin{center}
\includegraphics[width=18cm]{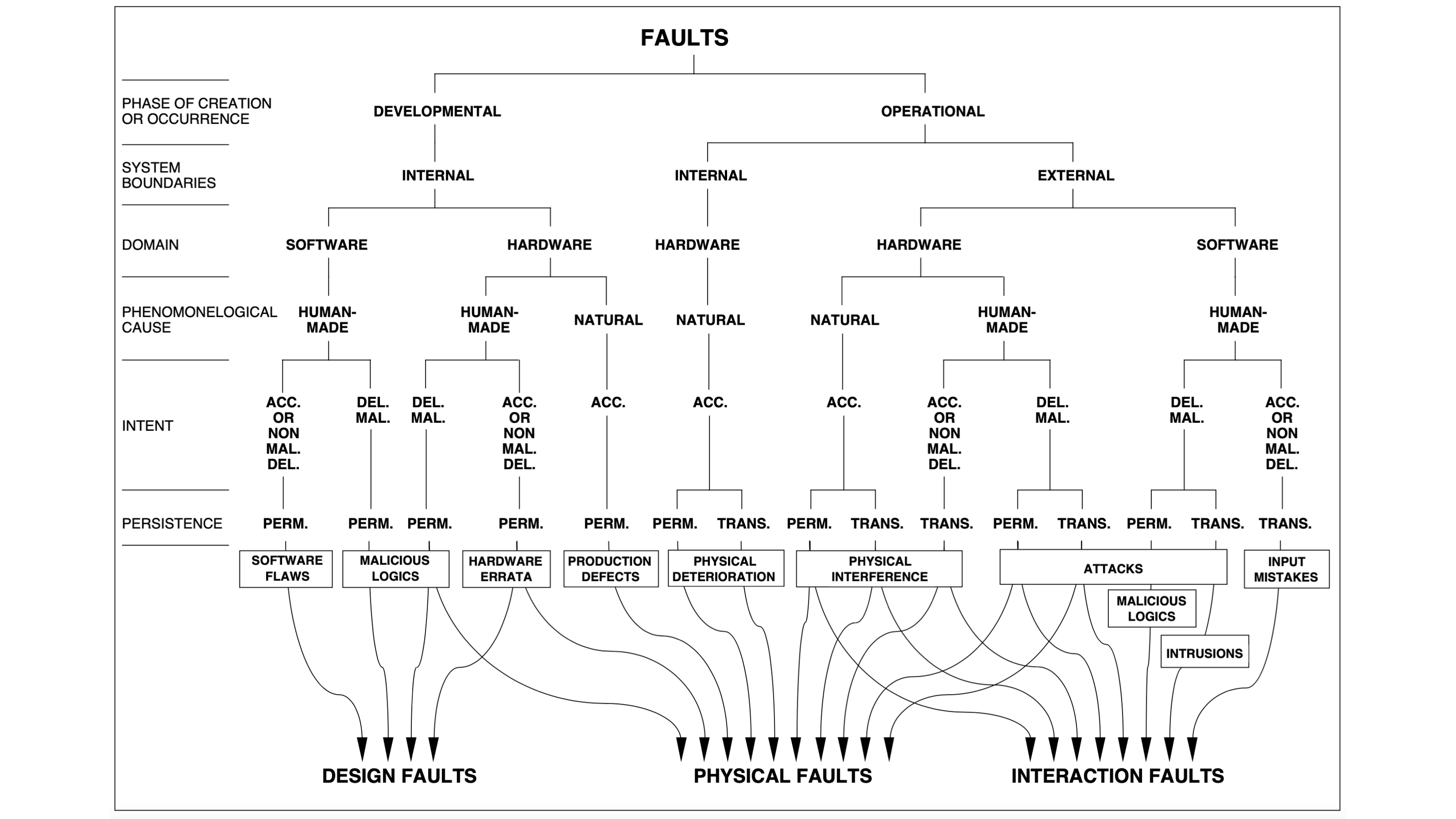}
\end{center}
\caption{Combined fault class diagram used in classical dependability and resilience analysis. Reproduced from \cite{avizienis2001fundamental}.}\label{fig:2}
\end{figure}

\begin{figure}[htp!]
\begin{center}
\includegraphics[width=\textwidth]{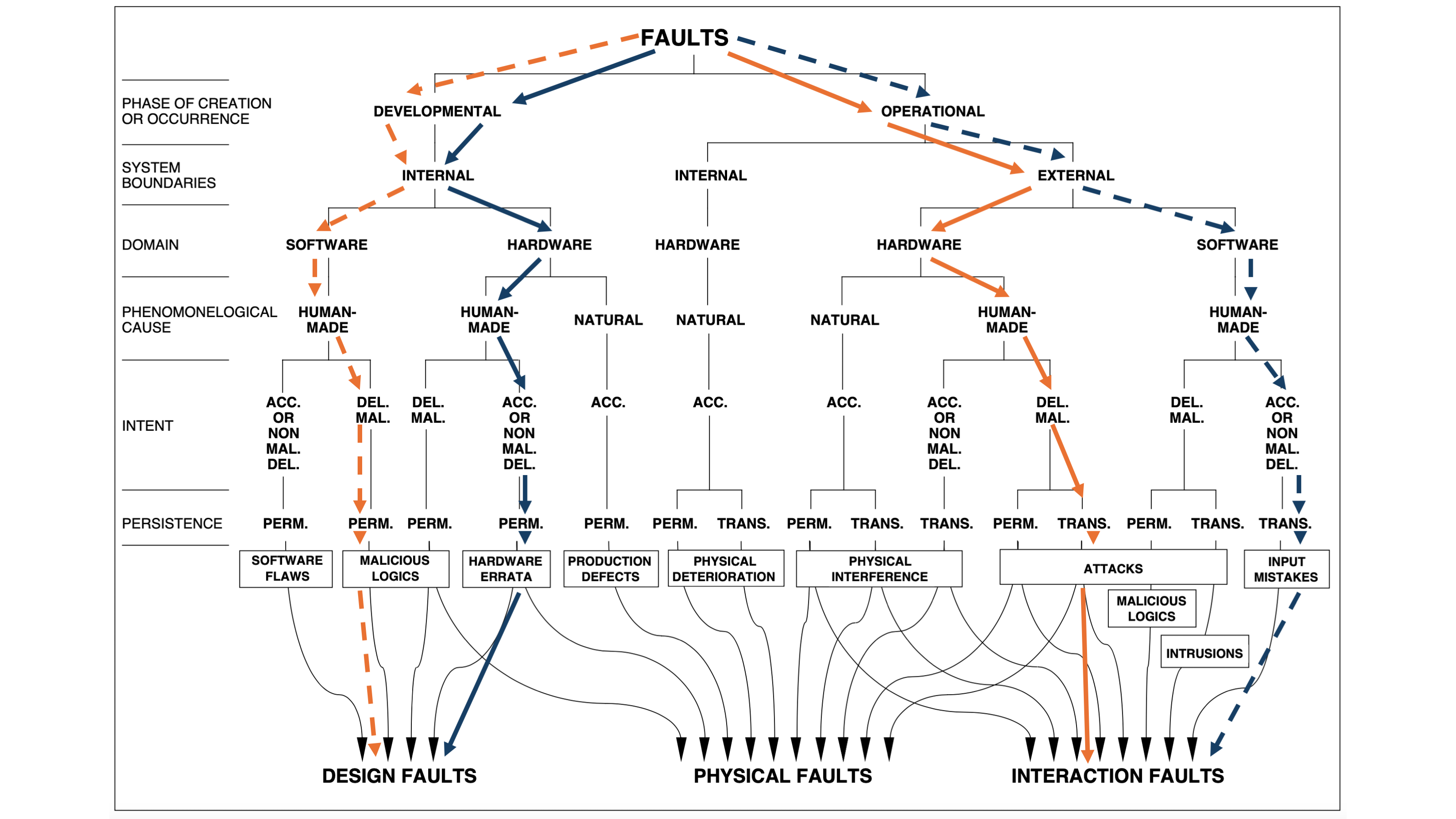}
\end{center}
\caption{Combined fault class diagram for stimuli challenging DCQCS resilience. Blue, dashed line: quantum testbed, miscalibrated qubits. Blue, solid line: quantum control ASIC microfabrication error. Orange, dashed line: malicious data snooping software embedded in an open source quantum compilation toolchain used by experimentalists. Orange, solid line: malicious noise injection in VQE execution using magnetic interference equipment. Adapted from \cite{giusto2025typology}.}\label{fig:3}
\end{figure}

Dependency and resilience analysis of classical computing systems \citep{avizienis2001fundamental} has produced classifications of events and stimuli with respect to their effects that can be directly applied to quantum computers and DCQCS. Take, for instance, combined fault class diagrams (CFCD) used to perform root cause analysis of faults (Figure \ref{fig:2}). Thanks to how general their definitions are, these afford us with the possibility of tracing root causes of risk extending to quantum stacks on both quantum testbeds and quantum vendor hardware \cite{giusto2025typology}. Figure \ref{fig:3} sketches four hypothetical cases along these lines.

\begin{itemize}
    \item \textbf{Quantum testbed, non-intentional (blue dashed line).} Manual qubit calibration performed in QISE laboratories is a laborious task, often required to elicit properties at the boundary of known physics. Without the assistance of automated means, the process can be imprecise, leading to unusable hardware and a subsequent arduous search for adequate physical parameters. Following the CFCD above, miscalibrations are operational faults of an external origin, enabled by software with a human operator that accidentally makes mistakes on the choice of input, which can be corrected by repeating the procedure. This case exemplifies an \textit{interaction fault}.
    \item \textbf{Quantum testbed, malicious (orange dashed line).} As testbeds mature, experiments performed in them go from simple calibration tasks to execution of circuits via quantum compilation. A multitude of quantum compilation toolchains have proliferated. We expect some of these tools to carry malicious code more frequently, intended to gather information for the purposes of geopolitically motivated academic espionage \citep{feder2019trade}, one of many possible motives. The latter corresponds to a developmental fault of internal origin that manifests in software, maliciously produced by human actors to have a lasting (almost permanent) effect to achieve program logic serving purposes different than enabling research. The above maps to a \textit{design fault}.
    \item \textbf{Quantum vendor system, non-intentional (blue solid line).} As qubit counts increase across available modalities, quantum control ASICs become inescapable \citep{castillo2023electronic,frank2023low}. Similarly to how classical computing hardware design and manufacturing experience errors, we expect to find flaws in their quantum counterparts as hardware companies and foundries adapt to the more aggressively stringent requirements of each component. These faults are this developmental, internal to the design and manufacturing process, solidified in hardware of accidental human origin that persist as hardware errata. Such a case maps to a \textit{design fault}.
    \item \textbf{Quantum vendor system, malicious (orange solid line).} We expect an increase in the deployment of quantum computing systems on-premise throughout this decade \citep{humble2021quantum,beck2024integrating}. These systems will become part of standard cyberinfrastructure facilities, which includes also acquiring their risks and associated mitigation strategies. VQE is a potential target due to its promising performance and computational properties across multiple domains and its susceptibility to noise at certain regimes. One may imagine an attack on a quantum resource that executes VQE through the placement of malicious hardware that produces noise profiles which can slow or disrupt valuable computations while masking its presence by acting somewhat randomly. This is a form of tampering \citep{ghosh2023primer}, and by extension an \emph{interaction fault}.
\end{itemize}

These seemingly distinct cases highlight the unifying explanatory power of existing methods and tools for dependability and resilience analysis. We note that the process of classifying faults satisfies requirements for \textit{reliability} analysis when assuming that a DCQCS starts from a normal (i.e., correct) state. How likely are these when quantum computing hardware is introduced?

Consider the case of bit-flip memory errors in contemporary classical cyberinfrastructure cause by various events such as cosmic rays \cite{asorey2023calculation}. Across exascale systems using recent hardware specifications, single-bit error rates of $10^{-18}$ and transistor counts of $10^{17}$ are typical. Three main reasons appear to be responsible for these scales of operation and integration. First, transistor technology became a highly optimized constructive unit capable of reliably rendering ideas in classical computer organization. Second, boolean and sequential logic depend exclusively on naturally bistable states. Third, the cost of full recovery for a single processing unit is low and tends to be relatively fast for hardware failures that do not require component replacement (e.g. performing a reset). It is also worth noting that, in time, parallelism and system design strategies have taken over transistor performance gains as the main source of improvement \cite{mcnairy2018exascale}.

In contrast, a DCQCS is a highly heterogeneous system with coupled analog-digital properties, no standardized constructive unit as of yet, and desirable error rates below $10^{-4}$, with a notable recent exception of $10^{-7}$ \citep{smith2024single}. The cost to achieve full recovery tends to be high in testbeds due to manual work, but has started to decrease for quantum vendor hardware thanks to ASIC development for calibration and optimal quantum control algorithms \cite{miller2025low, potovcnik2025scale}. In this sense, since QPU hardware requires a continuous-time description, the process of recovery from non-irreversible failures is closer to \textit{steering} than \textit{resetting}. By extension, assessing resilience requires introducing effective models to assist us in our quest for systems that can be justifiably trusted.

\subsection{Resilience has multiple fundamental explanatory frameworks}

A useful way to formalize the operation of a QPU from a purely physical perspective is through the Fokker-Planck equations \citep{risken1996fokker}. In general, for the time-dependent Schr\"odinger equation

\begin{equation}
    i \hbar \partial_t \psi(x_i,t) = -\frac{\hbar^2}{2m} \partial_{x_i}^2 \psi(x_i,t) + V(x_i,t)\psi(x_i,t)
\end{equation}

its corresponding Fokker-Planck equation reads

\begin{equation}
    \label{schfpe}
    \partial_t \psi(x_i,t) = \frac{i}{\hbar} \frac{\hbar^2}{2m} \partial_{x_i}^2 \psi(x_i,t) -\frac{i}{\hbar} V(x_i,t)\psi(x_i,t)
\end{equation}

For any observable $\mathcal{O}$, empirical knowledge of the system in terms of $\langle \psi | \mathcal{O}| \psi \rangle$ is restricted by the hardware's realizable sampling frequency due to the dynamics of quantum measurement. In a very concrete sense, the acquisition of information is restricted by an energy budget driven by qubit reset costs and circuit restart costs. Taking Eq. \ref{schfpe} into its more abstract restatement for the probability mass function $f$

\begin{equation}
    \partial_t f(x_i,t) = -\partial_{x_i} [\mu(x_i,t) f(x_i,t)] + \partial_{x_i}^2[D(x_i,t) f(x_i,t)]
\end{equation}

confirms that using rigorous quantum hardware simulations to understand the resilience of QPEs is feasible only for small-scale systems, particularly due to the stochastic integration techniques required by the $\partial_{x_i}^2 D(x,y)$ term. We can now imagine how modeling a DCQCS can proceed and why it is a hard task that is almost on par with constructing an actual system. Since it exemplifies a \emph{de facto} complex multiscale stochastic system \citep{nunez2024generalized}, we can imagine a collection of master equations that phenomenologically describe each level of abstraction of the system from the perspective of information flows and their enabling mechanisms (i.e., \textit{ enabling robotics}). Since QPUs Eq. \ref{schfpe} intrinsically contains the corresponding Lindblad equation describing the QPU \citep{de2023quantum}, we can in principle thus produce a hierarchy of coupled master equations that approximate the full behavior of the quantum stack in terms of data, programs and control signals. At this stage, the couplings between this stack of master equations in the QPU and the larger stack of master equations for HPC systems can be made. Curiously, the view described here and the view of quantum programming proposed by \cite{di2024abstraction} seem to match, as the interface between levels of the quantum stack may equate (or connect) to the formal description of their respective couplings from the viewpoint of master equations. The cost, however, is equivalent to building a digital twin of a QPU and the signals and effects propagating from it.

\begin{figure}[htp!]
\begin{center}
\includegraphics[width=0.8\textwidth]{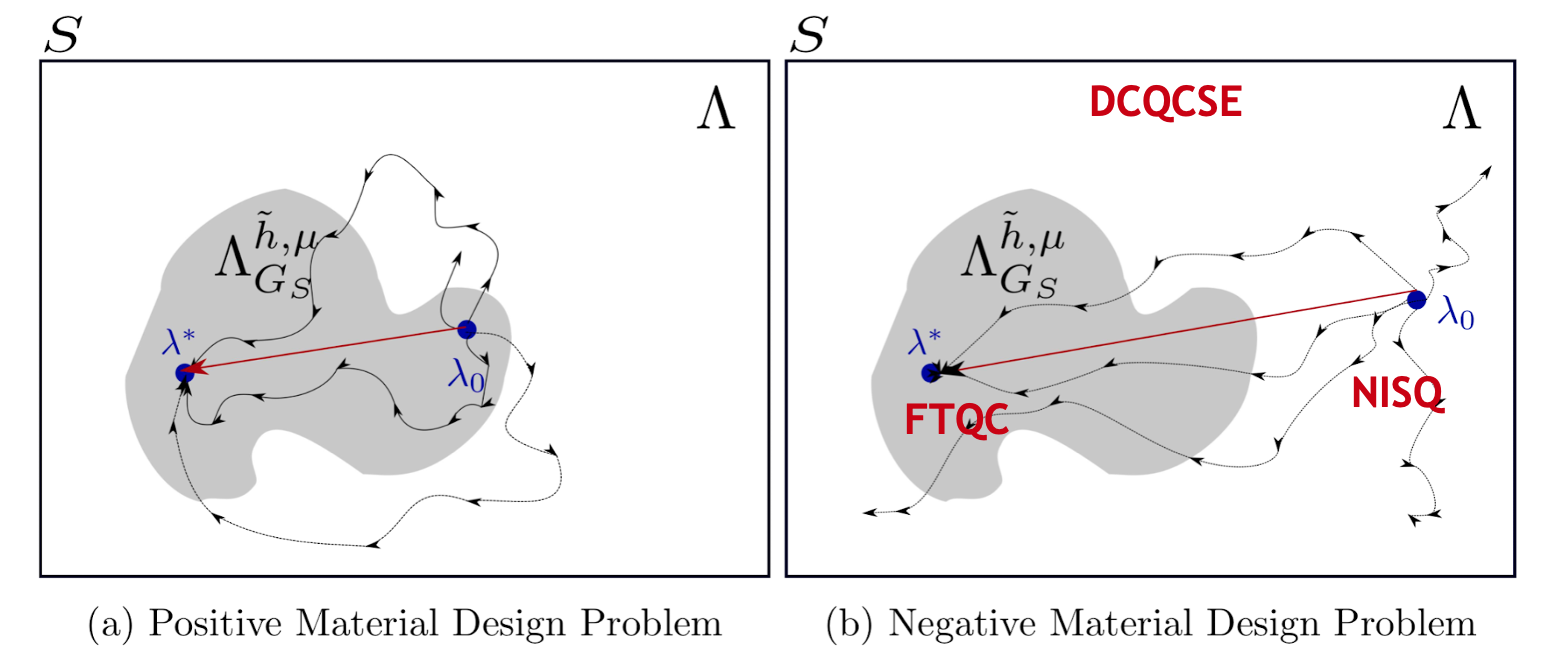}
\end{center}
\caption{Dependable Classical-Quantum Computer Systems Engineering as a negative material design problem. The difficulty of obtaining a DCQCS can be explained at large by the need to traverse the space of governing laws of coupled classical-quantum systems with limited orienting features and a starting non-solution. Adapted from \cite{nunez2020toward}.}\label{fig:4}
\end{figure}

Tn a very concrete sense, building DCQCS is no different from a \emph{material design problem}, or a problem of finding adequate governing laws that fully specify the structure and function of a system \cite{nunez2020toward}. By extension, building an architecture of modular and yet hierarchical control systems that ensure resilience under a multitude of stimuli is an even harder problem. We call here the corresponding overarching endeavor \emph{Dependable Classical-Quantum Computer Systems Engineering}. Although we have just started to achieve HPC-QPU integration with NISQ devices \citep{beck2024integrating,elsharkawy2024integration}, the properties, constraints, and opportunities imposed by fault-tolerant quantum computers (FTQC) are most likely not representative of the set of dynamical laws (i.e., the master equations governing each abstraction layer and its couplings) for future utility-scale devices (Figure \ref{fig:4}). The search for these laws is in general largely unstructured, and we start with a non-solution (i.e., NISQ-based DCQCS). In a very concrete sense, a holistic first-principles simulation strategy for DCQCSE may not be an intellectually economic route to resilience, but it provides crucial insights into the phenomenology of the integration of classical and quantum technologies.

\begin{figure}[h!]
\begin{center}
\includegraphics[width=0.6\textwidth]{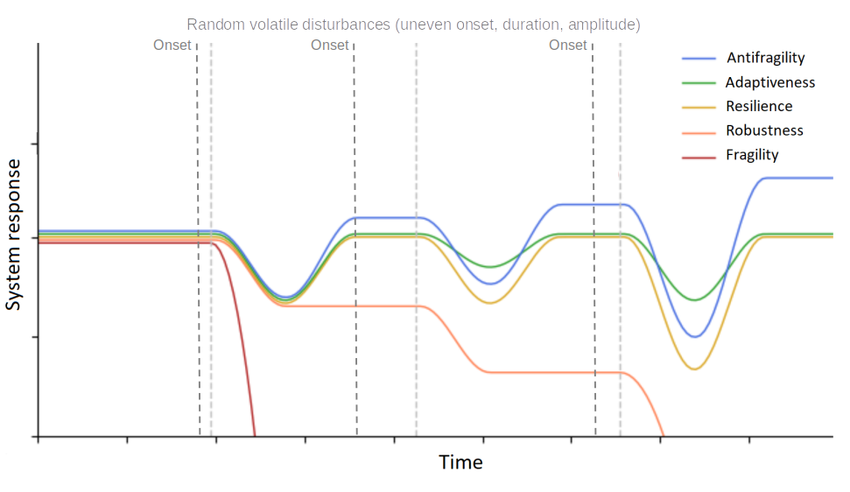}
\end{center}
\caption{Classes of systems from the perspective of the time-dependent response of dynamical systems to various kinds of perturbations. Reproduced from \cite{axenie2023antifragility}.}\label{fig:5}
\end{figure}

At a more conceptual level, we are interested in the dynamics enabling the design and interrogation of computer systems in a prescriptive manner. \emph{Prescriptive} here refers to the distinction between desirable and undesirable properties under a formal specification of correct functionality and expected performance. We suggest here that borrowing from dynamical systems theory is, once again, a productive strategy. In this case, we want to qualify the time-dependent response of a system as a function of the magnitude and classes of perturbations it can experience \citep{axenie2023antifragility}. \emph{Fragility}, the propensity of a system to be disrupted, constitutes a useful metric to choose (Figure \ref{fig:5}). We end this section by briefly mapping integration scales to known classical-quantum systems fragility states.

Physical qubits are intrinsically \textit{fragile} under various noise sources. One layer above, optimal quantum control provides \textit{robustness} in the form of improved T1 and T2 values up to device and experimental capabilities \citep{wilhelm2020introduction}. As advances in quantum hardware and control pile up and error thresholds give way to quantum error correction \cite{zimboras2025myths}, logical qubit \textit{resilience} provided by fault-tolerance percolates upwards into \textit{robustness} at the classical-quantum systems level. Building up on this, DCQCSE aims to achieve full-system \textit{resilience} by design \citep{giusto2024dependable}, which may include providing quantum simulators of various sorts (e.g., \cite{li2024tanq}) to endow a DCQCS with limited \textit{adaptiveness}. Neither classical nor quantum computer systems displaying \textit{antifragility}\footnote{One may speculate that antifragile QPUs are intrinsically not possible, since noise appears to be an irrecoverable perturbation in all cases. However, feedback loops and couplings to classical infrastructure may result in special antifragile states for DCQCS we are yet unaware of. For a detailed mathematical treatment of antifragility see \cite{taleb2013mathematical}.} are known today.

\section{Quantum Resilience Assessment for DCQCS}

We wish to suggest here that advancing the state of the art in DCQCS engineering toward resilience can be achieved by focusing on the systematic prevention, mitigation and study of faults leading to varying degrees of system-wide fragility that decrease end-user value as long as we develop theory to predict and explain cascades of compositional effects across hardware and software elements of the stack. Based on the discussion in the prior section, existing tools for computer dependability analysis are necessary but not sufficient to achieve resilience given the continuous-time, metastable properties that drive the hardware substrate in QPUs. By extension, adequate tools must harness observable properties of the corresponding dynamical systems and their couplings to produce probabilistically bounded predictions which can be either simulated or corroborated experimentally, either of each in an efficient manner. Resilience assessment methods developed by the civil engineering community appear to fit the requirements of such task due to the direct transferability of the concepts and principles to classical-quantum computer systems infrastructure.

\begin{figure}[h!]
\begin{center}
\includegraphics[width=0.8\textwidth]{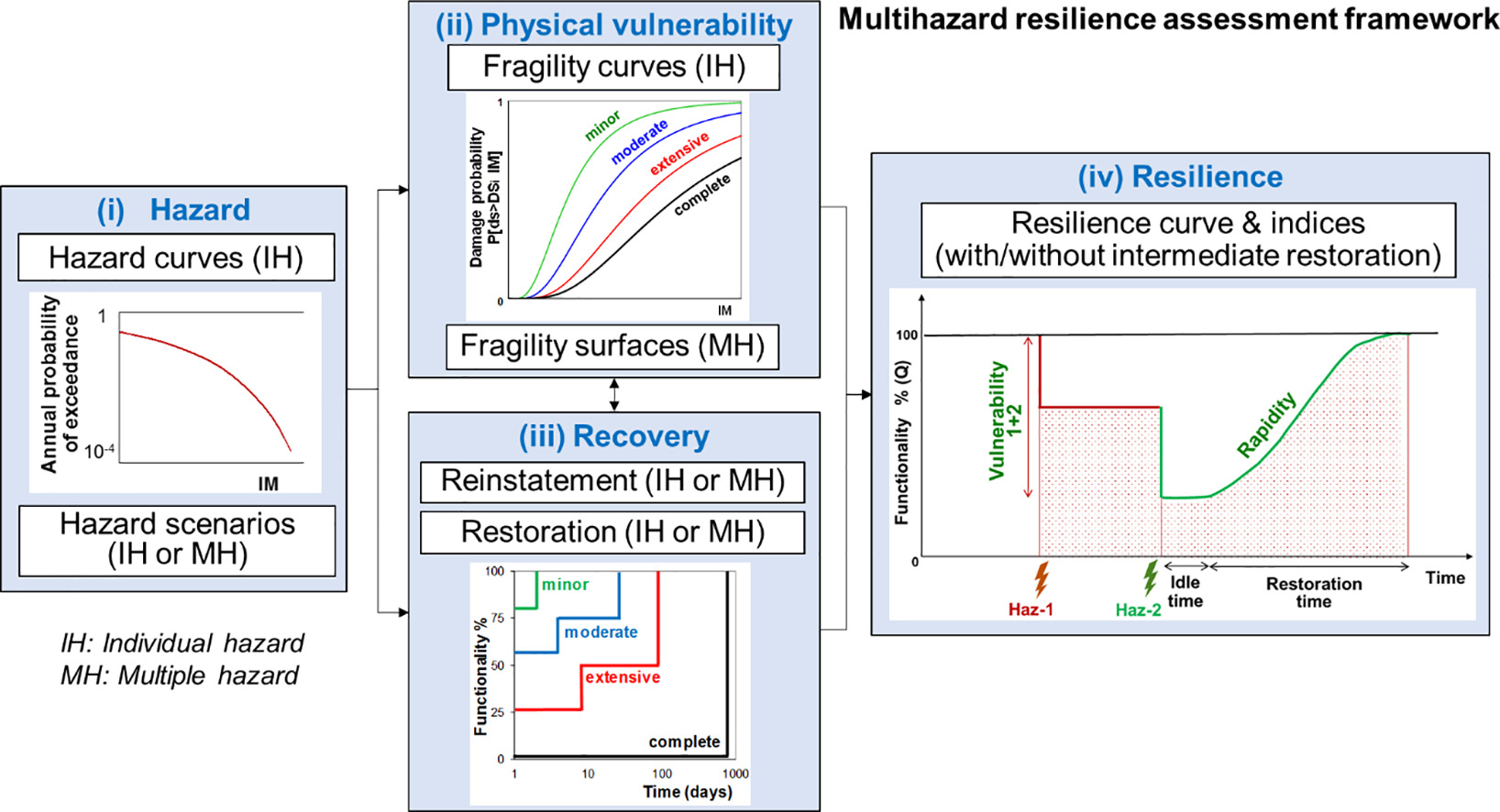}
\end{center}
\caption{Simplified diagram of the multihazard resilience assessment framework commonly used in civil engineering to evaluate infrastructure. Reprinted from \cite{argyroudis2020resilience}.}\label{fig:6}
\end{figure}

Resilience assessment is a vast subject, and only a cursory summary is given here. A bird's eye definition would characterize it as the scientific root-cause imputation processes involving fragilities, hazards, and risks for specific infrastructure instances \citep{porter2015beginner}. It is simultaneously a process of theory-driven estimation and evidence-driven reasoning: ground truths leading to hazard and risk estimates need to arise from characterizable properties of the substrate responsible for the structure and function of a system. The purpose of this process is to make a decision about how to allocate resources intended to minimize potential future loss in that system during its design or operation or to choose among various recovery strategies that restore functionality to full or partial functionality. Figure \ref{fig:6} conceptually depicts the process described above. We attempt a preliminary reduction from some of the main concepts in resilience assessment in civil engineering to DCQCS engineering as follows. 

\subsection{Resiliency Assessment applies to dependable HPC-QPU integration}

Classical and quantum processing units constitute \textit{assets} characterizable in terms of \textit{attributes}. Assets can be subject to \textit{environmental excitations} such as external noise sources, user input, operation errors or malicious attacks. \textit{Hazard} defines the relationship between degree of environmental excitation and \textit{exceedance probability} (probability of exceeding threshold values of relevance); for instance, depolarizing noise rates exceeding $10^{-3}$ per qubit. \textit{Fragility} describes a function that relates probability of occurrence of an undesirable outcome such as dephasing noise causing qubit errors to fall below the QEC threshold (i.e., a \textit{damage state}) to a measure of demand such as the intensity of the dephasing noise. A \textit{failure} then occurs when a hazard impacts an asset to produce a specific damage state of a certain magnitude and exceedance probability. We define \emph{loss} as a metric for an undesirable outcome, such as a superconducting full-chip failure or a spurious resonator response; with it, \emph{risk} becomes a relation between the degree of undesirable outcome and its exceedance probability. \textit{Vulnerability}, in contrast, describes the relationship between the degree of an undesirable outcome to the intensity of a demand on an asset. For instance, the probability of permanent single-qubit damage as a certain voltage threshold is exceeded constitutes a vulnerability.

The general form of the resilience assessment framework adequately matches the needs of DCQCS. Exposure data must inform the analysis process, which motivates structured experimentation and reuse driven by asset categories (e.g., control systems, laboratory instrumentation, ASIC types, qubit modalities). During asset analysis, multiple theoretical descriptions of classical and quantum systems provide response models to the necessary excitations. Quantum optics generally provides an abstract description of quantum systems \citep{browne2017quantum}, while circuit quantum electrodynamics applies directly to superconducting quantum platforms \citep{blais2021circuit}. Resilience assessment often requires having default models when specific ones are not available. Later, hazards are quantified in terms of the types and levels of excitation severity versus the probability of exceeding each of these levels. With these tools, loss analysis quantifies how diminished functionality or negative consequences impact various value-generating attributes of assets. As an example, consider how different noise sources may slow the convergence of VQE. The end of the process includes models for how to perform recovery: a quantification of how long it would take to reinstate functionality after a hazard materializes, and to restore the system after a given damage state has been reached. We observe here that establishing these practices must be rooted in showing how effort spent in them translates into user value, an discussion we defer to the next section of this article.

We find four final advantages of resilience assessment. The first one is its ability to integrate multiple concerns simultaneously, since it is intended for complex infrastructure from the start. DCQCS comprise a multitude of components with different dynamics and descriptions, which makes it a hard target already. Second, loss models can be chained across hierarchies of systems to naturally reflect the propagation of consequences throughout coupled degrees of freedom; this feature helps us construct a progressively causal view of faults, errors and failures through compositional reasoning and generative effects \citep{adam2017systems}. Third, the frameworks is self-refining, which implies that a resilience assessment produced by imperfect versions of fragility, vulnerability and loss models is preferable to no information at all; this is even more so for DCQCS at their infancy in which systematic guidance needs to be constructed. Fourth and last, the process is systematic, which as demonstrated in a specific effort for civil engineering \citep{van2023interdependent}, lends itself naturally to community building and data sharing.

\subsection{Three QRA case study sketches}

The following short examples illustrate how QRA would operate under three putative scenarios in a hypothetical superconducting multi-qubit testbed.

\begin{figure}[h!]
\begin{center}
\includegraphics[width=0.8\textwidth]{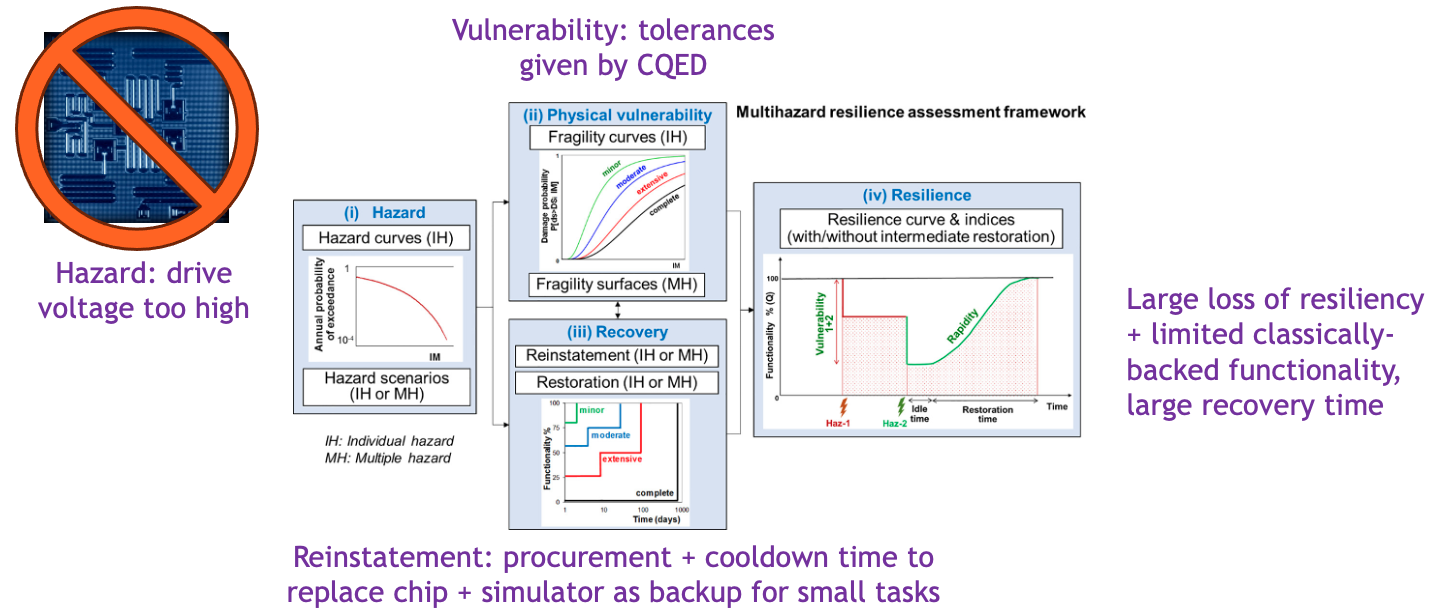}
\end{center}
\caption{QRA elements for a full-chip failure in a superconducting qubit tesbed.}\label{fig:7}
\end{figure}

\textbf{Sketch 1: full chip damage.} Consider a situation within an academic testbed in which careless manipulation of drive voltages exceeded electrical thresholds with respect to the design of the superconducting chip (Figure \ref{fig:7}). Hazard-wise, one may associate a low probability of occurrence (e.g., serious misconfiguration by an incoming graduate student) with a high-impact event. Knowing the parameters of the device and data from prior or focused experimental runs, it would be possible to have estimates for the corresponding fragility curves at various voltage intensities. If it materializes, the recovery time is likely bound by procurement time, and can be expected to be long in relation to routine events in the system. Restoration of service, for small tasks, may involve simulators that can partially support scientific discover processes.

\begin{figure}[h!]
\begin{center}
\includegraphics[width=0.8\textwidth]{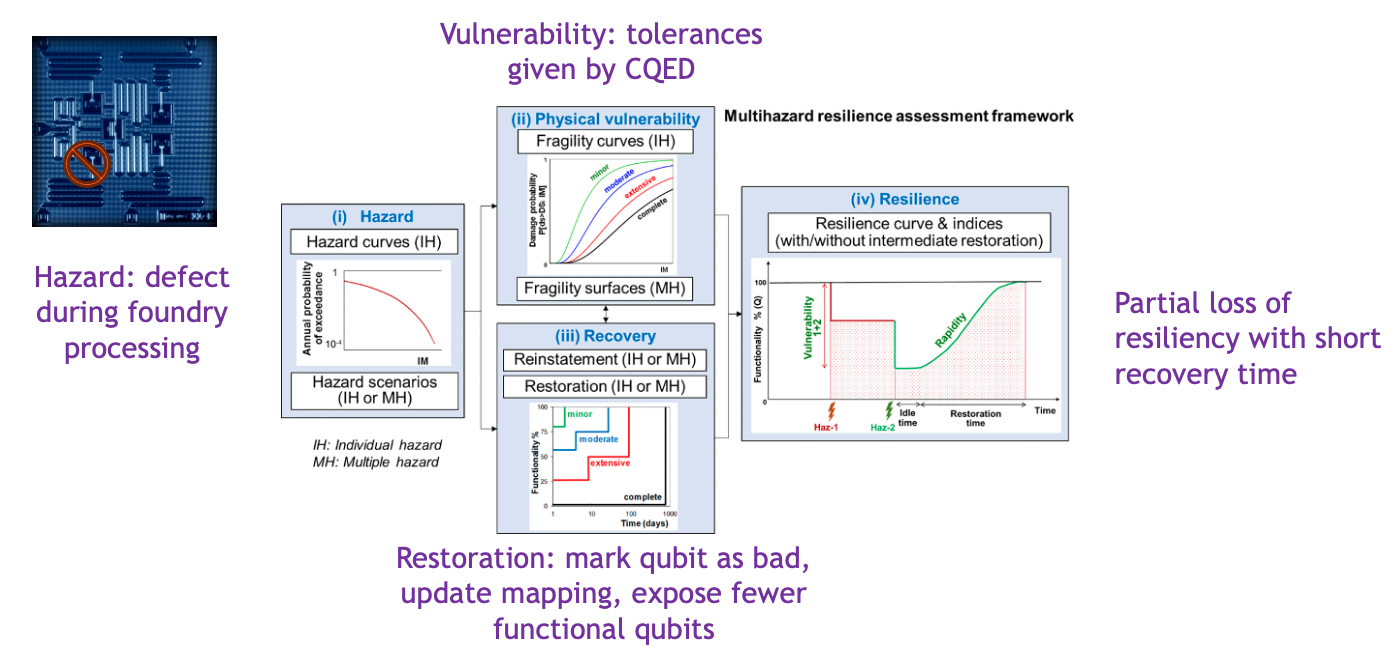}
\end{center}
\caption{QRA elements for a single qubit failure in a superconducting qubit tesbed.}\label{fig:8}
\end{figure}

\textbf{Sketch 2: single-qubit damage.} Now, instead of human error in the lab, a defect in one of the superconducting chips renders a qubit unusable (Figure \ref{fig:8}) --e.g., spurious resonator response in close proximity to a Josephson junction \citep{rafferty2021spurious}. As experience increases with a given fabrication facility, it may be possible to characterize this response and use circuit QED to extract information for preliminary fragility curves. Recovery entails here marking the qubit as unusable and updating the corresponding mapping. This leads to partial loss of resilience with a short recovery time below complete functionality.

\begin{figure}[h!]
\begin{center}
\includegraphics[width=0.8\textwidth]{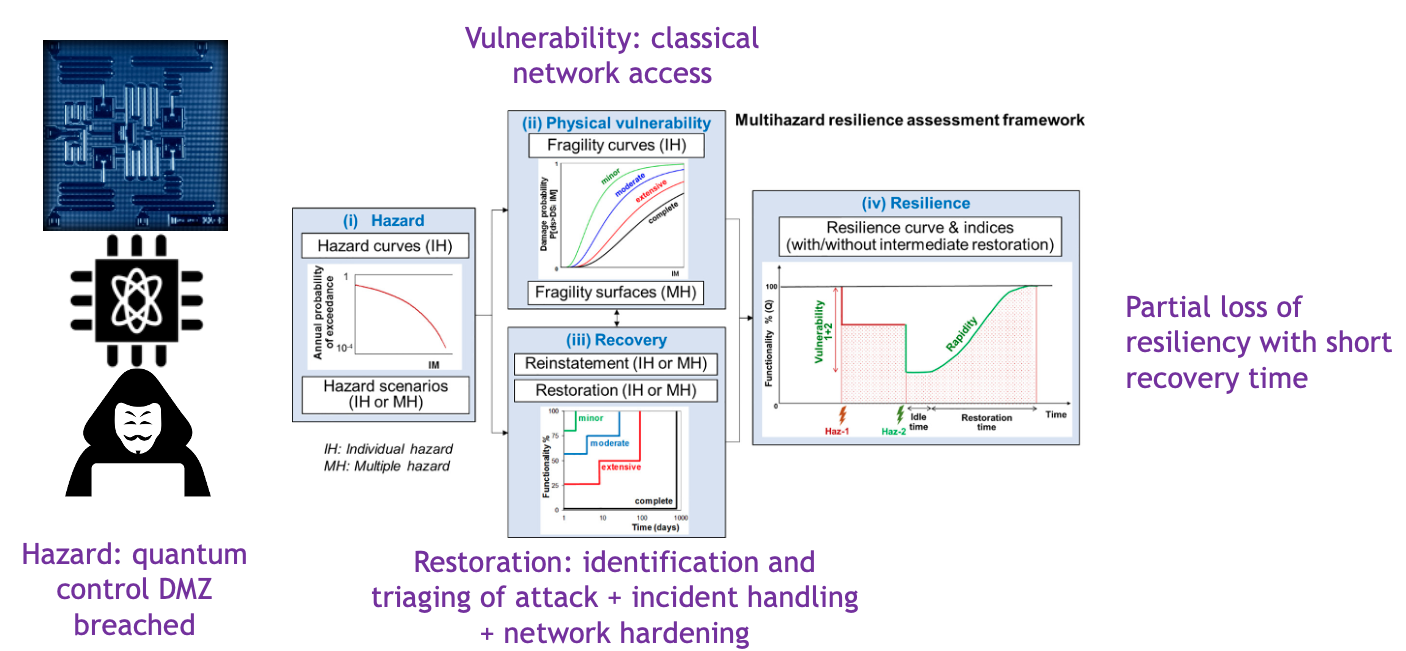}
\end{center}
\caption{QRA elements for a malicious attack on quantum control hardware through a DMZ network.}\label{fig:9}
\end{figure}

\textbf{Sketch 3: malicious access to quantum control hardware} Finally, let us now suppose a malicious actor capable of gaining unauthorized access to quantum control hardware through the laboratory DMZ network (Figure \ref{fig:9}). Fortunately, such a threat tends to be version-specific for hardware, firmware and software versions, and its frequency of occurrence likely documented in cybersecurity communities. Both hazard and fragilities become feasible to estimate from data. Recovery-wise, standard incident handling can help patch network vulnerabilities to harden the network and return to normalcy with full functionality in short time.

\section{Justifying the long term cost of QRE: user value of HPC-QPU resources}

Integrating the complex interrelated factors leading to successful Quantum Resiliency Assessment methods and tools in DCQCS will undoubtedly be an expensive endeavor. Pragmatically, the majority of this cost already transfers from hardware manufacturers and integrators to HPC centers and ultimately end-users through either cloud-based pricing schemes or facility access costs. Quantum hardware is in general more expensive to design, produce, and operate than its classical counterpart. In consequence, the cost of these activities in DCQCS at scale will likely be a non-linear function of the number of specific couplings required for both classes of systems to achieve user value. The larger the scale of a DCQCS, the more likely its cost will rise and, barring the amortizing effect of multiple systems being produced and operated in relatively large numbers, the more expensive user access becomes. We should expect this to remain so until we discover new broader and more efficient principles behind DCQCS scalability. What, then, justifies the cost of adding complexity \textit{now} to ensure the resilience of computing processes performed with DCQCS \textit{later}?

To answer the question above, we need to provide a working definition of end-user value of DCQCS cyberinfrastructure. \textit{End users} of DCQCS will mostly comprise domain experts that seek to harness quantum resources to solve problems repeatedly for which classical ones result in intractability. We differentiate \textit{value} from \textit{end-user satisfaction}, historically introduced to characterize the perception of the adequacy of alignment of the design of a system with the end user's intended purpose and mechanics of use \citep{rivard1988factors}. \emph{Value}, in this context, means a numerical quantification of direct benefits users obtain after inspecting the results of a computational task (e.g., financial gains or savings, number of articles produced, increases in typical problem size). Despite the fact that quantifying the value of solving a given problem may be a hard task to accomplish, we will proceed here assuming that it is somehow possible for the sake of the current argument. We postulate here that if the value enabled by a DCQCS is at least equal to some threshold of interest for a sufficient number of users, then long-term investments into Quantum Resiliency Assessment will be justified across life cycle of a system. Given sufficient evidence, we expect this process to become a virtuous, self-sustaining process capable of simultaneously delivering increasing end-user value and dependability.

As a necessary conceptual exercise, we construct here a simple model of value connected with QRA exercises, suitable for characterizing the impact of specific events on DCQCS cyberinfrastructure. In the simplest case, we expect the value $V$ of a DCQCS to be proportional to its throughput $T$, the number of tasks completed per unit of time, or

\begin{equation}
    V \propto T.
\end{equation}

More realistically, throughput by itself does not reliably capture value unless pertinent goals are achieved. The ultimate motive of a computation is to achieve an impact $I$; one may imagine quantum economic advantage \citep{bova2023quantum} as a reasonable proxy for it. We may extend the definition of value to

\begin{equation}
    V \propto T \cdot I
\end{equation}

The sheer differences observed in the expected computational complexity of known quantum algorithms \citep{dalzell2023quantum} and the most frequent problems \citep{arnault2024typology} they solve suggest a more granular model in which throughput and impact depend on the type of task $j \in J$. The corresponding modification becomes

\begin{equation}
    V \propto \sum_j T_j \cdot I_j.
\end{equation}

Not all tasks of the same type are created equal. Impact tends to be a function of the scale (i.e., instance size) of the problem being solved. In realistic situations, only instances above a given threshold start to yield benefits given the cost of operating a piece of cyberinfrastructure. We can therefore assume the existence of a threshold function $\alpha(s_{i,j})$ that modulates impact, with probability $p(s_{i,j})$ of observing instance size $s_{i,j}$. Thus, the new value estimate is

\begin{equation}
    V \propto \sum_i \sum_j p(s_{i,j}) \cdot T_j \cdot I_j^{\alpha(s_{i,j})}.
\end{equation}

A similar argument can be made about the \textit{quality} of the solutions obtained. By quality we mean proximity to an exact solution. In the case of quantum algorithms, quality becomes proximity to the expected distributions of computations performed were ideal (noiseless) qubits available. Lower quality, at a high level, can be explained by imperfections in technology (e.g. coherent control errors \citep{greiwe2023effects,berberich2024robustness}) or sensitivity to inevitable noise sources \citep{fontana2021evaluating,xue2021effects,bharti2022noisy}, both abundant at this point in the transition from NISQ to FTQC \citep{liang2024modeling}; benchmarking plays a substantial role in this regard \citep{resch2021benchmarking}. All this allows us to hypothesize the existence of a non-negative function $\gamma(s_{i,j})$ over instance sizes per each task that further modulates impact estimates. We also expect $\gamma$ to be inversely proportional to an error function $\epsilon(s_{i,j})$ stated in terms of imperfections and noise sources such that $\epsilon = \alpha^{-1}$. Consequently,

\begin{align}
    V & \propto \sum_i \sum_j p(s_{i,j}) \cdot T_j \cdot I_j^{\alpha(s_{i,j}) \cdot \gamma(s_{i,j})} \nonumber \\ 
      & \propto \sum_i \sum_j p(s_{i,j}) \cdot T_j \cdot I_j^{\alpha(s_{i,j})/\epsilon(s_{i,j})}
\end{align}

Finally, addition of QPU hardware to HPC infrastructure is motivated by the need to have repeated use to solve different problem instances, usually under time constraints. Solutions are not indefinitely valuable in time, especially in mission-critical or industrial contexts. This is clearly a case of computational rationality \citep{gershman2015computational}: the value of HPC-QPU integration lies not in its implied technical feats, but in how it reduces the impedance mismatch between user needs with concrete, specific time constraints, and limited classical hardware capabilities. Therefore, the value in DCQCS varies dynamically and assumes that after a certain variable characteristic time $\tau_{i.j}$, specific per task and instance size, the value of a solution will decay rapidly as a function $d(\tau_{i,j} - t(s_{i,j}))$. Note that $\mathcal{T}(t) = (\tau_{i,j})$ specifies a time-dependent matrix that is externally defined by the community of users, as well as per-task input-size probabilities $\mathcal{P}(t) = (p(s_{i,j}))$. Our value estimation equation becomes

\begin{equation}
    V(\mathcal{T}(t), \mathcal{P}(t)) \propto \sum_i \sum_j p(s_{i,j}) \cdot d(\tau_{i,j} - t(s_{i,j})) \cdot T_j \cdot I_j^{\alpha(s_{i,j})/\epsilon(s_{i,j})}.
\end{equation}

We are now in a position to qualify how the resiliency of DCQCS impacts user value term by term using this simplistic model. Throughput $T_j$ suffers from faults leading to service interruption or system performance degradation. A system which no longer runs tasks produces no value, and a system capable of processing fewer tasks reduces overall value proportionally. Faults that decrease performance, whether due to longer single-task completion times or longer queuing times, also risk a steeper penalty in the form of greater decay of per-task impact as per $t(s_{i,j})$. Impact is indirectly impaired in two different ways. Faults may disrupt the number of qubits available without necessarily increasing execution time or decreasing throughput. Depending on qubit modality, faults with greater consequences will likely bias usage away from use cases most sensitive to them, hence altering the distribution of $p(s_{i,j})$ and creating artificial constraints that limit the generality of a multi-user system.

In these situations, particularly until utility-scale fault-tolerant quantum computers become available, relatively small qubit counts induce a heightened price per qubit lost, thus limiting the ability to solve problems large enough to demonstrate utility or advantage sustainably with existing resources as given by $\alpha(s_{i,j})$. Faults that compromise qubit and entanglement fidelities increase error rates captured by $\epsilon(s_{i,j})$ and, by extension, decrease the quality of the output. Observe that both $\alpha$ and $\epsilon$ modulate $I_j$ non-linearly, making empirical benchmarking and use case testing fundamental for resiliency analysis to understand their impact. Although $\mathcal{T}(t)$ and $\mathcal{P}(t)$ are supplied externally (although not necessarily explicitly), repeated encounters of users with unrecoverable events, decreased system throughput for groups of tasks, or decreased performance per class of tasks can lead to selective loss of trust in DCQCS capabilities. 

\section{Conclusions}

Throughout this article, we have laid out an argument in favor of a new, emergent set of engineering practices to advance the state of classical-quantum computer systems (DCQCS) toward dependability. Resilience is central to delivering end-user value in the form of algorithmic progress and application performance in classical computing; we expect this to remain so as quantum resources are introduced. At the same time, the complexity behind the design, manufacture, deployment, and operation of QPUs appears to be a few orders of magnitude higher than that for CPU, GPU and other technologies. Consequently, one of the contributions of this article is an attempt to clarify and unify the language used to describe our engineering aspirations for DCQCS while pointing to theoretical and empirical tools to satisfy them. It is also clear that accounting quantitatively for the effect of faults on value generation for end users can transparently justify investing in Quantum Resilience Assessment (QRA) methods of the sort described in the prior section. QRA will be undoubtedly perceived as unnecessary expensive as long as its role in accelerating the development of utility-scale fault-tolerant quantum computing systems remains opaque. We state here that QRA will contribute to the advent of utility-scale FTQC; even more, we state that achieving it will necessarily involve this or other similar methodologies and frameworks.

Ernest Rutherford allegedly divided science into ``\textit{either physics or stamp collecting}'' \citep{bernal1939social}. Research on resiliency in DCQCS is in a state in which both ``physics'' and ``stamp collecting'' are needed. Regarding the ``physics'' part of the equation in DCQCS, we have already suggested a research program to gain greater understanding around the dynamics of quantum and classical dynamical systems and their coupling, their translation into manifestations of fragility, and the need to understand the limits of rigorous QPU simulation as devices scale up: theoretical challenges in HPC-QPU integration abound and remain in need of serious work. At the same time, we require collecting evidence systematically and collectively on QPUs both standalone and co-located with HPC resources. From standardized data formats to collect contributed data from quantum and classical-quantum devices and their exposures, to benchmarking QPUs under the associated operational regimes they impose, and automating inference processes as new DCQCS scale in size and complexity, more and better evidence is needed. Quantum Resilience Assessment as a framework preliminary appears to serve as the glue that combines physics and stamp collecting into a new synthesis of classical-quantum computing architecture.

At a technical level, our discussion points to two potentially interesting and productive research problems. One of them is complementing QRA frameworks with agent-based modeling in lieu of real HPC-QPU systems: these can simulate complex systems in which value is a well-defined notion and human interventions can occur \citep{heydari2018guiding}, and ABM frameworks can be accelerated by contemporary HPC resources \citep{leyba2024simcov} to make simulations feasible. Contrary to rigorous simulations of quantum devices, ABM models would focus on modeling the interfaces defined by information exchange across layers of the classical-quantum stack with sufficient representative of relevant states across control, data and program backplanes. The second point is on the potential of quantum analog computers to model the fine-grained dynamics of digital quantum systems feasibly. As mentioned before, classical rigorous stochastic simulations of devices governed by quantum optics rapidly become infeasible as these scale. This would result in a strong limitation for QRA when estimating hazard and fragility curves of specific quantum hardware if data about them is scarce. Recent literature \citep{shao2024rydberg} suggests that certain quantum analog architectures efficiently simulate quantum optics, which would help accelerate the theoretical and computational characterization of QPUs intended for DCQCS integration. Both of these potential opportunities demonstrate a need for tools to allow us to efficiently (and systematically) reason about and design tests for quantum computing technologies altogether.

The conceptual and speculative tone across this work precisely mimics how early we are in the HPC-QPU integration race, and by extension how much we have not yet discovered of the design space they span. What we know, however, is that computing systems that have withstood the test of time share three common attributes: they deliver measurable value to their users, they are programmable in intellectually efficient ways, and \textit{they can be justifiably trusted by their users}. This last property entails the need for resilience as part of DCQCS, which calls for broad cooperation across relevant disciplines and the opportunity to accelerate the prospect of utility-scale fault-tolerant devices capable of interacting with classical HPC in meaningful and valuable ways for users. While much more work is needed to further the claims presented here, indirect and preliminary practical experience along these lines appears to favor the potential impact of such a research program.

\section*{Acknowledgments}
The author thanks the National Center for Supercomputing Applications, the Illinois Center for Quantum Information Science and Technology and the Program for Arms Control \& Domestic and International Security at the University of Illinois Urbana-Champaign, as well as Edoardo Giusto (University of Naples Frederick II) and Iskandar Sitdikov (IBM Quantum) for motivating the writing of these ideas. Finally, special thanks go to Travis Humble and Yuhong Song (Oak Ridge National Laboratory) whose request for a talk on this topic at StableQ 2024 helped solidify the skeleton of the argument presented here. Conversations with Laura Schultz (Argonne National Laboratory) have challenged productively many of the ideas contained here. This work was partially funded by the Leadership-Class Compute Facility project (NSF \#2323116) and the QLCI Hybrid Quantum Architectures and Networks (NSF \#2016136).

\bibliographystyle{unsrt}  
\bibliography{references}

\begin{thebibliography}{10}

\bibitem{berberich2024quantum}
Julian Berberich and Daniel Fink.
\newblock Quantum computing through the lens of control: A tutorial introduction.
\newblock {\em IEEE Control Systems}, 44(6):24--49, 2024.

\bibitem{butko2020understanding}
Anastasiia Butko, George Michelogiannakis, Samuel Williams, Costin Iancu, David Donofrio, John Shalf, Jonathan Carter, and Irfan Siddiqi.
\newblock Understanding quantum control processor capabilities and limitations through circuit characterization.
\newblock In {\em 2020 International Conference on Rebooting Computing (ICRC)}, pages 66--75. IEEE, 2020.

\bibitem{chohan2024comparative}
Akash Chohan.
\newblock A comparative review of quantum bits: Superconducting, topological, spin, and emerging qubit technologies, 2024.

\bibitem{gambetta2022quantum}
Jay Gambetta.
\newblock Quantum-centric supercomputing: The next wave of computing.
\newblock {\em IBM Research Blog}, 2022.

\bibitem{aaronson2009need}
Scott Aaronson and Andris Ambainis.
\newblock The need for structure in quantum speedups.
\newblock {\em arXiv preprint arXiv:0911.0996}, 2009.

\bibitem{arnault2024typology}
Pablo Arnault, Pablo Arrighi, Steven Herbert, Evi Kasnetsi, and Tianyi Li.
\newblock A typology of quantum algorithms.
\newblock {\em arXiv preprint arXiv:2407.05178}, 2024.

\bibitem{beck2024integrating}
Thomas Beck, Alessandro Baroni, Ryan Bennink, Gilles Buchs, Eduardo Antonio~Coello P{\'e}rez, Markus Eisenbach, Rafael~Ferreira da~Silva, Muralikrishnan~Gopalakrishnan Meena, Kalyan Gottiparthi, Peter Groszkowski, et~al.
\newblock Integrating quantum computing resources into scientific hpc ecosystems.
\newblock {\em Future Generation Computer Systems}, 161:11--25, 2024.

\bibitem{elsharkawy2024integration}
Amr Elsharkawy, Xiaorang Guo, and Martin Schulz.
\newblock Integration of quantum accelerators into hpc: Toward a unified quantum platform.
\newblock {\em arXiv preprint arXiv:2407.18527}, 2024.

\bibitem{giusto2024dependable}
Edoardo Giusto, Santiago Nu{\~n}ez-Corrales, Phuong Cao, Alessandro Cilardo, Ravishankar~K Iyer, Weiwen Jiang, Paolo Rech, Flavio Vella, Bartolomeo Montrucchio, Samudra Dasgupta, et~al.
\newblock Dependable classical-quantum computer systems engineering.
\newblock {\em arXiv preprint arXiv:2408.10484}, 2024.

\bibitem{avizienis2001fundamental}
Algirdas Avizienis, Jean-Claude Laprie, Brian Randell, et~al.
\newblock Fundamental concepts of dependability.
\newblock {\em Technical Report Series-University of Newcastle upon Tyne Computing Science}, 2001.

\bibitem{nielsen2006optimal}
Michael~A Nielsen, Mark~R Dowling, Mile Gu, and Andrew~C Doherty.
\newblock Optimal control, geometry, and quantum computing.
\newblock {\em Physical Review A—Atomic, Molecular, and Optical Physics}, 73(6):062323, 2006.

\bibitem{boscain2021introduction}
Ugo Boscain, Mario Sigalotti, and Dominique Sugny.
\newblock Introduction to the pontryagin maximum principle for quantum optimal control.
\newblock {\em PRX Quantum}, 2(3):030203, 2021.

\bibitem{koch2022quantum}
Christiane~P Koch, Ugo Boscain, Tommaso Calarco, Gunther Dirr, Stefan Filipp, Steffen~J Glaser, Ronnie Kosloff, Simone Montangero, Thomas Schulte-Herbr{\"u}ggen, Dominique Sugny, et~al.
\newblock Quantum optimal control in quantum technologies. strategic report on current status, visions and goals for research in europe.
\newblock {\em EPJ Quantum Technology}, 9(1):19, 2022.

\bibitem{nunez2023quantum}
Santiago N{\'u}{\~n}ez-Corrales.
\newblock Quantum abstract machines without circuits: the need for higher algorithmic expressiveness.
\newblock {\em arXiv preprint arXiv:2307.08422}, 2023.

\bibitem{di2024abstraction}
Olivia Di~Matteo, Santiago N{\'u}{\~n}ez-Corrales, Micha{\l} Stech{\l}y, Steven~P Reinhardt, and Tim Mattson.
\newblock An abstraction hierarchy toward productive quantum programming.
\newblock {\em arXiv preprint arXiv:2405.13918}, 2024.

\bibitem{klir1991requisite}
George~J Klir and W~Ross Ashby.
\newblock Requisite variety and its implications for the control of complex systems.
\newblock {\em Facets of systems science}, pages 405--417, 1991.

\bibitem{rallis2025interfacing}
Konstantinos Rallis, Ioannis Liliopoulos, Georgios~D Varsamis, Evangelos Tsipas, Ioannis~G Karafyllidis, Georgios~Ch Sirakoulis, and Panagiotis Dimitrakis.
\newblock Interfacing quantum computing systems with high-performance computing systems: an overview.
\newblock {\em arXiv preprint arXiv:2509.06205}, 2025.

\bibitem{burgholzer2026munich}
Lukas Burgholzer, Jorge Echavarria, Patrick Hopf, Yannick Stade, Damian Rovara, Ludwig Schmid, Erc{\"u}ment Kaya, Burak Mete, Muhammad~Nufail Farooqi, Minh Chung, et~al.
\newblock The munich quantum software stack: Connecting end users, integrating diverse quantum technologies, accelerating hpc.
\newblock In {\em Proceedings of the Supercomputing Asia and International Conference on High Performance Computing in Asia Pacific Region}, pages 55--67, 2026.

\bibitem{shehata2025building}
Amir Shehata, Peter Groszkowski, Thomas Naughton, Murali Gopalakrishnan~Meena, Elaine Wong, Daniel Claudino, Rafael Ferreira~da Silvaa, and Thomas Beck.
\newblock Building a software stack for quantum-hpc integration.
\newblock {\em arXiv e-prints}, pages arXiv--2503, 2025.

\bibitem{ahmed2025telemetry}
Hossam Ahmed, Burak Mete, Helmut Heller, Matthew Tovey, Xiaolong Deng, Asim Zulfiqar, Muhammad~Nufail Farooqi, Mahmoud Abuzayed, Martin Schulz, and Laura Schulz.
\newblock Telemetry for quantum systems in hpc centers.
\newblock In {\em ISC High Performance 2025 Research Paper Proceedings (40th International Conference)}, pages 1--11. Prometeus GmbH, 2025.

\bibitem{giusto2025typology}
Edoardo Giusto, Santiago N{\'u}{\~n}cz-Corrales, Alessandro Cilardo, Nicola Mazzocca, and Travis Humble.
\newblock A typology of quantum-classical faults.
\newblock In {\em 2025 55th Annual IEEE/IFIP International Conference on Dependable Systems and Networks Workshops (DSN-W)}, pages 188--195. IEEE, 2025.

\bibitem{feder2019trade}
Toni Feder.
\newblock Trade wars and other geopolitical tensions strain us--china scientific collaborations.
\newblock {\em Physics Today}, 72(11):22--26, 2019.

\bibitem{castillo2023electronic}
Stefanie Castillo.
\newblock The electronic control system of a trapped-ion quantum processor: A systematic literature review.
\newblock {\em IEEE Access}, 11:65775--65786, 2023.

\bibitem{frank2023low}
David~J Frank, Sudipto Chakraborty, Kevin Tien, Pat Rosno, Mark Yeck, Joseph~A Glick, Raphael Robertazzi, Ray Richetta, John~F Bulzacchelli, Daniel Ramirez, et~al.
\newblock Low power cryogenic rf asics for quantum computing.
\newblock In {\em 2023 IEEE Custom Integrated Circuits Conference (CICC)}, pages 1--8. IEEE, 2023.

\bibitem{humble2021quantum}
Travis~S Humble, Alexander McCaskey, Dmitry~I Lyakh, Meenambika Gowrishankar, Albert Frisch, and Thomas Monz.
\newblock Quantum computers for high-performance computing.
\newblock {\em IEEE Micro}, 41(5):15--23, 2021.

\bibitem{ghosh2023primer}
Swaroop Ghosh, Suryansh Upadhyay, and Abdullah~Ash Saki.
\newblock A primer on security of quantum computing.
\newblock {\em arXiv preprint arXiv:2305.02505}, 2023.

\bibitem{asorey2023calculation}
Hern{\'a}n Asorey and Rafael Mayo-Garcia.
\newblock Calculation of the high-energy neutron flux for anticipating errors and recovery techniques in exascale supercomputer centres.
\newblock {\em The Journal of Supercomputing}, 79(8):8205--8235, 2023.

\bibitem{mcnairy2018exascale}
Cameron McNairy.
\newblock Exascale fault tolerance challenge and approaches.
\newblock In {\em 2018 IEEE International Reliability Physics Symposium (IRPS)}, pages 3C--4. IEEE, 2018.

\bibitem{smith2024single}
MC~Smith, AD~Leu, K~Miyanishi, MF~Gely, and DM~Lucas.
\newblock Single-qubit gates with errors at the $10^{-7}$ level.
\newblock {\em arXiv preprint arXiv:2412.04421}, 2024.

\bibitem{miller2025low}
Nathan~Eli Miller, Laith~A Shamieh, and Saibal Mukhopadhyay.
\newblock Low-latency digital feedback for stochastic quantum calibration using cryogenic cmos.
\newblock In {\em 2025 Design, Automation \& Test in Europe Conference (DATE)}, pages 1--7. IEEE, 2025.

\bibitem{potovcnik2025scale}
Anton Poto{\v{c}}nik.
\newblock How to scale the electronic control systems of a quantum computer.
\newblock {\em Nature Electronics}, 8(1):3--4, 2025.

\bibitem{risken1996fokker}
Hannes Risken.
\newblock {Fokker-Planck equation for several variables; methods of solution}.
\newblock In {\em The Fokker-Planck Equation: Methods of Solution and Applications}, pages 133--162. Springer, 1996.

\bibitem{nunez2024generalized}
Santiago N{\'u}{\~n}ez-Corrales and Eric Jakobsson.
\newblock A generalized theory of interactions--i. the general problem.
\newblock {\em arXiv preprint arXiv:2403.02346}, 2024.

\bibitem{de2023quantum}
M{\'a}rio~J de~Oliveira.
\newblock {Quantum Fokker-Planck structure of the Lindblad equation}.
\newblock {\em Brazilian Journal of Physics}, 53(5):121, 2023.

\bibitem{nunez2020toward}
Santiago Nunez-Corrales.
\newblock {\em Toward a unified view of complex multiscale stochastic systems: a generalized theory of interactions and its computational infrastructure for their universal and efficient investigation}.
\newblock PhD thesis, University of Illinois at Urbana-Champaign, 2020.

\bibitem{axenie2023antifragility}
Cristian Axenie, Oliver Lopez-Corona, Michail~A Makridis, Meisam Akbarzadeh, Matteo Saveriano, Alexandru Stancu, and Jeffrey West.
\newblock Antifragility as a complex system’s response to perturbations, volatility, and time.
\newblock {\em ArXiv}, 2023.

\bibitem{wilhelm2020introduction}
Frank~K Wilhelm, Susanna Kirchhoff, Shai Machnes, Nicolas Wittler, and Dominique Sugny.
\newblock An introduction into optimal control for quantum technologies.
\newblock {\em arXiv preprint arXiv:2003.10132}, 2020.

\bibitem{zimboras2025myths}
Zolt{\'a}n Zimbor{\'a}s, B{\'a}lint Koczor, Zo{\"e} Holmes, Elsi-Mari Borrelli, Andr{\'a}s Gily{\'e}n, Hsin-Yuan Huang, Zhenyu Cai, Antonio Ac{\'\i}n, Leandro Aolita, Leonardo Banchi, et~al.
\newblock Myths around quantum computation before full fault tolerance: What no-go theorems rule out and what they don't.
\newblock {\em arXiv preprint arXiv:2501.05694}, 2025.

\bibitem{li2024tanq}
Ang Li, Chenxu Liu, Samuel Stein, In-Saeng Suh, Muqing Zheng, Meng Wang, Yue Shi, Bo~Fang, Martin Roetteler, and Travis Humble.
\newblock Tanq-sim: Tensorcore accelerated noisy quantum system simulation via qir on perlmutter hpc.
\newblock {\em arXiv preprint arXiv:2404.13184}, 2024.

\bibitem{taleb2013mathematical}
Nassim~Nicholas Taleb and Raphael Douady.
\newblock Mathematical definition, mapping, and detection of (anti) fragility.
\newblock {\em Quantitative Finance}, 13(11):1677--1689, 2013.

\bibitem{porter2015beginner}
Keith Porter.
\newblock A beginner’s guide to fragility, vulnerability, and risk.
\newblock {\em Encyclopedia of earthquake engineering}, 2015:235--260, 2015.

\bibitem{argyroudis2020resilience}
Sotirios~A Argyroudis, Stergios~A Mitoulis, Lorenzo Hofer, Mariano~Angelo Zanini, Enrico Tubaldi, and Dan~M Frangopol.
\newblock Resilience assessment framework for critical infrastructure in a multi-hazard environment: Case study on transport assets.
\newblock {\em Science of the Total Environment}, 714:136854, 2020.

\bibitem{browne2017quantum}
Dan Browne, Sougato Bose, Florian Mintert, and MS~Kim.
\newblock From quantum optics to quantum technologies.
\newblock {\em Progress in Quantum Electronics}, 54:2--18, 2017.

\bibitem{blais2021circuit}
Alexandre Blais, Arne~L Grimsmo, Steven~M Girvin, and Andreas Wallraff.
\newblock Circuit quantum electrodynamics.
\newblock {\em Reviews of Modern Physics}, 93(2):025005, 2021.

\bibitem{adam2017systems}
Elie~M Adam.
\newblock {\em Systems, generativity and interactional effects}.
\newblock PhD thesis, Massachusetts Institute of Technology, 2017.

\bibitem{van2023interdependent}
John~W van~de Lindt, Jamie Kruse, Daniel~T Cox, Paolo Gardoni, Jong~Sung Lee, Jamie Padgett, Therese~P McAllister, Andre Barbosa, Harvey Cutler, Shannon Van~Zandt, et~al.
\newblock The interdependent networked community resilience modeling environment (in-core).
\newblock {\em Resilient Cities and Structures}, 2(2):57--66, 2023.

\bibitem{rafferty2021spurious}
O~Rafferty, S~Patel, CH~Liu, S~Abdullah, CD~Wilen, DC~Harrison, and R~McDermott.
\newblock Spurious antenna modes of the transmon qubit.
\newblock {\em arXiv preprint arXiv:2103.06803}, 2021.

\bibitem{rivard1988factors}
Suzanne Rivard and Sid~L Huff.
\newblock Factors of success for end-user computing.
\newblock {\em Communications of the ACM}, 31(5):552--561, 1988.

\bibitem{bova2023quantum}
Francesco Bova, Avi Goldfarb, and Roger~G Melko.
\newblock Quantum economic advantage.
\newblock {\em Management science}, 69(2):1116--1126, 2023.

\bibitem{dalzell2023quantum}
Alexander~M Dalzell, Sam McArdle, Mario Berta, Przemyslaw Bienias, Chi-Fang Chen, Andr{\'a}s Gily{\'e}n, Connor~T Hann, Michael~J Kastoryano, Emil~T Khabiboulline, Aleksander Kubica, et~al.
\newblock Quantum algorithms: A survey of applications and end-to-end complexities.
\newblock {\em arXiv preprint arXiv:2310.03011}, 2023.

\bibitem{greiwe2023effects}
Felix Greiwe, Tom Kr{\"u}ger, and Wolfgang Mauerer.
\newblock Effects of imperfections on quantum algorithms: A software engineering perspective.
\newblock In {\em 2023 IEEE International Conference on Quantum Software (QSW)}, pages 31--42. IEEE, 2023.

\bibitem{berberich2024robustness}
Julian Berberich, Daniel Fink, and Christian Holm.
\newblock Robustness of quantum algorithms against coherent control errors.
\newblock {\em Physical Review A}, 109(1):012417, 2024.

\bibitem{fontana2021evaluating}
Enrico Fontana, Nathan Fitzpatrick, David~Mu{\~n}oz Ramo, Ross Duncan, and Ivan Rungger.
\newblock Evaluating the noise resilience of variational quantum algorithms.
\newblock {\em Physical Review A}, 104(2):022403, 2021.

\bibitem{xue2021effects}
Cheng Xue, Zhao-Yun Chen, Yu-Chun Wu, and Guo-Ping Guo.
\newblock Effects of quantum noise on quantum approximate optimization algorithm.
\newblock {\em Chinese Physics Letters}, 38(3):030302, 2021.

\bibitem{bharti2022noisy}
Kishor Bharti, Alba Cervera-Lierta, Thi~Ha Kyaw, Tobias Haug, Sumner Alperin-Lea, Abhinav Anand, Matthias Degroote, Hermanni Heimonen, Jakob~S Kottmann, Tim Menke, et~al.
\newblock Noisy intermediate-scale quantum algorithms.
\newblock {\em Reviews of Modern Physics}, 94(1):015004, 2022.

\bibitem{liang2024modeling}
Qiyao Liang, Yiqing Zhou, Archismita Dalal, and Peter Johnson.
\newblock Modeling the performance of early fault-tolerant quantum algorithms.
\newblock {\em Physical Review Research}, 6(2):023118, 2024.

\bibitem{resch2021benchmarking}
Salonik Resch and Ulya~R Karpuzcu.
\newblock Benchmarking quantum computers and the impact of quantum noise.
\newblock {\em ACM Computing Surveys (CSUR)}, 54(7):1--35, 2021.

\bibitem{gershman2015computational}
Samuel~J Gershman, Eric~J Horvitz, and Joshua~B Tenenbaum.
\newblock Computational rationality: A converging paradigm for intelligence in brains, minds, and machines.
\newblock {\em Science}, 349(6245):273--278, 2015.

\bibitem{bernal1939social}
John~D Bernal.
\newblock {\em The Social Function of Science}.
\newblock Cambridge/MIT Press, 1939.

\bibitem{heydari2018guiding}
Babak Heydari and Michael~J Pennock.
\newblock Guiding the behavior of sociotechnical systems: The role of agent-based modeling.
\newblock {\em Systems Engineering}, 21(3):210--226, 2018.

\bibitem{leyba2024simcov}
Kirtus Leyba, Steven Hofmeyr, Stephanie Forrest, Judy Cannon, and Melanie Moses.
\newblock Simcov-gpu: Accelerating an agent-based model for exascale.
\newblock In {\em Proceedings of the 33rd International Symposium on High-Performance Parallel and Distributed Computing}, pages 322--333, 2024.

\bibitem{shao2024rydberg}
Xiao-Qiang Shao, Shi-Lei Su, Lin Li, Rejish Nath, Jin-Hui Wu, and Weibin Li.
\newblock Rydberg superatoms: An artificial quantum system for quantum information processing and quantum optics.
\newblock {\em Applied Physics Reviews}, 11(3), 2024.

\end{thebibliography}

\end{document}